\documentclass[fleqn,10pt]{wlscirep}
\usepackage{color}
\usepackage{graphicx}
\usepackage{dcolumn}
\usepackage{bm}
\usepackage{hyperref}
\usepackage{subfig}
\usepackage{epstopdf}
\usepackage{color}
\usepackage{amsmath}

\newcommand{\onlinecite}[1]{\hspace{-1 ex} \nocite{#1}\citenum{#1}}

\begin{document}


\title{\emph{Ab Initio} Approach to Second-order Resonant Raman Scattering Including Exciton-Phonon Interaction}
\author[1,3,*]{Yannick Gillet}
\author[2,3]{Stefan Kontur}
\author[1,3]{Matteo Giantomassi}
\author[2,3]{Claudia Draxl}
\author[1,3]{Xavier Gonze}

\affil[1]{Universit\'e catholique de Louvain, 
 Institute of Condensed Matter and Nanosciences, 
 Nanoscopic Physics,
 Chemin des \'etoiles 8, bte L7.03.01, 
 1348 Louvain-la-Neuve, Belgium
}%
\affil[2]{%
 Physics Department and IRIS Adlershof, 
 Humboldt-Universit\"at zu Berlin, 
 Zum Gro{\ss}en Windkanal 6, 
 D-12489 Berlin, Germany
}%
\affil[3]{%
 European Theoretical Spectroscopy Facility (ETSF)
}%
\affil[*]{yannick.jp.gillet@gmail.com}

\date{\today}

\begin{abstract}
Raman spectra obtained by the inelastic scattering of light by crystalline solids
contain contributions from first-order vibrational processes (e.g. the emission 
or absorption of one phonon, a quantum of vibration)
as well as higher-order processes with at least two phonons being involved. 
At second order, coupling with the 
entire phonon spectrum induces a response that may strongly depend on 
the excitation energy, and reflects complex
processes more difficult to interpret. 
In particular, excitons (i.e. bound electron-hole 
pairs) may enhance the absorption and emission of light, and couple strongly 
with phonons in resonance conditions.
We design and implement a first-principles methodology to compute 
second-order Raman scattering, incorporating dielectric responses and phonon 
eigenstates obtained from density-functional theory and many-body theory. We 
demonstrate our approach for the case of silicon, relating frequency-dependent 
relative Raman intensities, that are in excellent agreement with experiment, to 
different vibrations and regions of the Brillouin zone. We show that 
exciton-phonon coupling, computed from first principles, indeed strongly affect the spectrum in resonance 
conditions.
The ability to analyze second-order Raman spectra thus provides direct insight into this interaction.

\end{abstract}

\maketitle

\section*{Introduction}
In Raman spectroscopy, the change of frequency and intensity of
light, that is inelastically scattered by the material, allows one to 
collect a rich set of information, even more if the effect is monitored as a function
of the frequency of the incoming light~\cite{Yu2010,Weber2000,Cardona1975,Cardona1982}. In most systems, the energy loss is 
due to the creation of one or more phonons, and different orders of phonon-photon 
scattering processes contribute to the scattering cross section. The first-order 
contribution, in which only one phonon is emitted or absorbed, yields information on 
vibrations with very low crystalline momentum, due to the negligible momentum of 
light. Extracting information on phonon frequencies from 
the first-order spectrum is straightforward, as clear selection rules 
govern the appearance or extinction of usually well-isolated 
peaks. In contrast, second-order Raman processes exhibit features 
coming from different crystalline momenta, potentially from the entire Brillouin zone (BZ), 
with the only constraint of negligible momentum of the phonon pairs
involved in the two-phonon process.
They are more complex and, hence, more difficult 
to interpret.
When the frequency 
of the incoming light comes close to the specific frequency needed 
to drive the transfer of an electron from an occupied state to an 
unoccupied state, which is termed a resonance, the absolute Raman 
intensities can change by several orders of magnitude. 
This resonance phenomenon appears independently of the number of emitted phonons.
Thus a particular frequency can be selected to increase the Raman signal~\cite{Cardona1975}. 
The possibility to create bound electron-hole pairs, called excitons, crucially 
impact optical absorption and related photocurrent in many materials, a mechanism at work in 
commercial photovoltaic devices~\cite{Lanzani2012,Pan2015,Akselrod2014,Nozik2012}.
Understanding and characterizing their dynamics and 
interactions is an experimental challenge. In this context, the ability 
to interpret the widely available Raman data 
to characterize exciton-phonon interaction is crucial
for the understanding of light-matter interaction in general,
and in particular in technologically relevant materials.

Experimentally, the second-order part of the Raman spectrum is 
often used to characterize low-dimensional or layered materials, such as e.g. graphene~\cite{Ferrari2013} 
or graphite~\cite{Thomsen2000}, and more recently also transition-metal 
dichalcogenides~\cite{Berkdemir2013,Corro2014,Gaur2015,Wang2015,Corro2016}. The few existing 
first-principles studies for these systems relied on the independent-electron approximation (IPA) 
\cite{Venezuela2011,Herziger2014}, in which the formation of excitons
is neglected. As these low-dimensional materials exhibit 
very well-defined and specific spectral features~\cite{Ferrari2004}, their 
second-order Raman spectra can still be interpreted quite easily,
at least concerning the location of the peaks. 
However, for photovoltaic applications, bulk materials form the vast majority of relevant materials.
Their Raman spectra contains a larger number of 
features, and is thus difficult to decipher. Resonant first-order and 
second-order Raman scattering processes may have e.g. different temperature behavior 
\cite{Weber2016}, whose most natural interpretation invoke characteristics of 
the direct and indirect gaps. For classical semiconductors such as silicon (and Ge, III-V etc),
the analysis of second-order Raman spectra was facilitated by the use of simple Hamiltonians, based on the effective mass approximation or on semi-empirical pseudopotentials, with model phonon band structure, model electron-phonon interaction and, in some cases, the inclusion of Wannier-Mott excitons~\cite{Go1975,Cardona1985,Trallero1989,Grein1991,Garcia1994}. Early first-principles calculations were done for vanishing laser frequencies (no resonance effects) without excitonic effects~\cite{Strauch1996}.
Our work allows one to compute from first-principles second-order Raman spectra without any of these limitations.

\begin{figure*}[hbt]
\centering
\includegraphics[width=16cm]{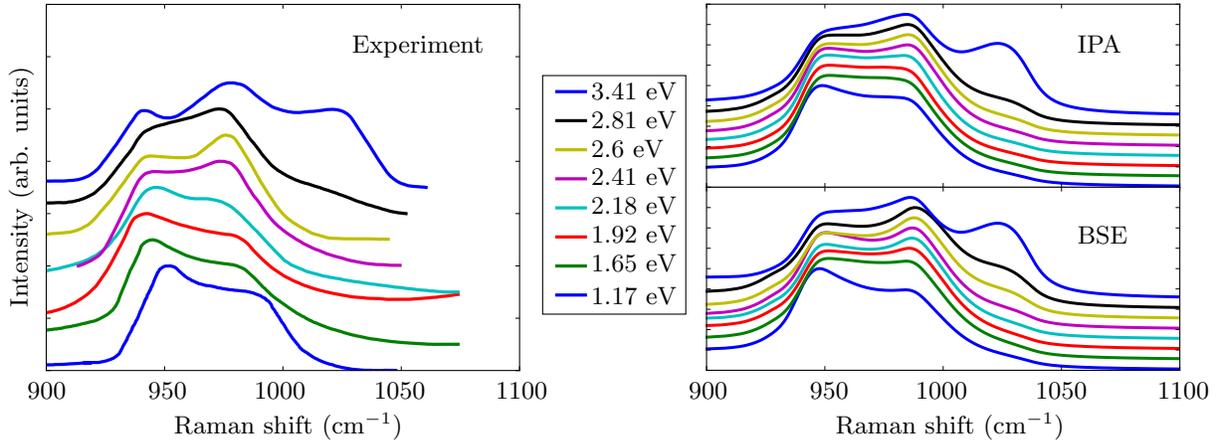}
\caption{(Color online). Experimental (left, taken from Ref. \onlinecite{Renucci1975}) 
and theoretical (right) second-order Raman spectra of silicon, between 900 and 1050 cm$^{-1}$, for different excitation energies between 1.17 eV and 3.41 eV, normalized 
to their maximal values. For clarity, the curves are 
shifted with respect to each other (lowest excitation energy at the bottom). 
Theoretical results have been obtained in the independent-particle approximation (IPA) and in the 
Bethe-Salpeter equation (BSE) formalism.}
\label{fig:renucci}
\end{figure*}

A prototypical example of resonant second-order Raman spectra is 
presented in Fig. \ref{fig:renucci}. It shows the experimental 
frequency-dependent 
Raman spectra of silicon between 900 and 1050 cm$^{-1}$, as 
measured by Renucci and coworkers~\cite{Renucci1975}. The spectra are characterized by 
significant variations in intensity depending on the frequency of the incoming light,
even for frequencies corresponding to energies smaller than the direct gap of 3.2 eV. Such 
resonance effects are allowed in second-order Raman processes, due to indirect 
transitions between electronic states that are caused by phonons of finite 
wavevector. Hence, it is clear that multi-phonon processes beyond first order 
are crucial. Below, we will show that all the features can be well interpreted 
by our first-principle theoretical approach.

A general methodology for describing multi-phonon scattering 
processes up to arbitrary order has been proposed by Knoll and Draxl 
\cite{Knoll1995}, but its application had been long restricted to the 
\emph{ab initio} calculations of first-order resonant Raman scattering,
in the IPA, without excitonic effects, e.g. in 
various 
superconducting materials \cite{Ambrosch-Draxl2002,Puschnig2001,Ravindran2003} 
and ladder compounds \cite{Spitaler2007,Spitaler2008,Spitaler2009}. 
Recently, thanks to the treatment of excitons using the
Bethe-Salpeter equation (BSE), excitonic effects were included
from first principles in such first-order resonant Raman scattering~\cite{Gillet2013}.
Excitonic effects were clearly observed in the absolute intensity enhancement, 
and proved crucial to obtain good agreement with experimental data. 
Other first-principles studies of Raman intensities have been based on 
density-functional perturbation theory~\cite{Baroni1986,Baroni1987,Gonze1995,Baroni2001,Veithen2004,
Veithen2005,Gonze2005a,Caracas2011}.
However, this last approach has been developed only in the limit of vanishing light 
frequency, so without addressing any resonance effect and without excitonic effects~\cite{Strauch1996}.

In this work, we address the even more challenging computation of
second-order resonant Raman scattering, including exciton-phonon 
coupling. Even by today standards, the computational load is 
quite formidable. We compute the ingredients using state-of-the-art methodology
(density-functional 
perturbation theory and Bether-Salpeter equation), while 
also showing the adequacy of the simplifying ``overtone'' 
approximation. 
Applying it to the case of Si, we not only reproduce the above described experimental data, 
but give insight into the origin of the spectral features, and explore the role of excitonic 
effects on the Raman intensities.

\section*{Results}

\begin{figure*}[hbt]
 \centering
 \includegraphics[width=8cm]{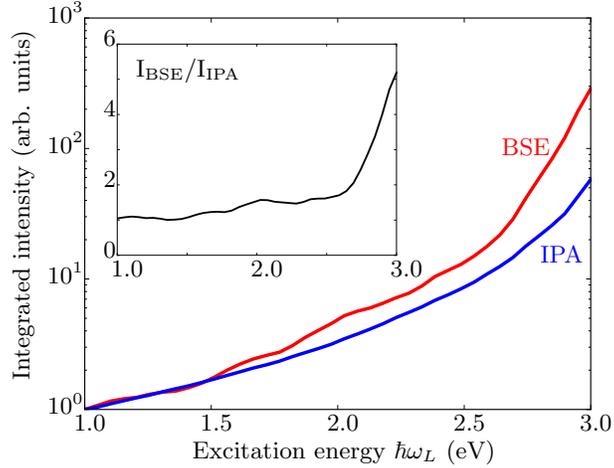}
 \caption{(Color online). Integrated intensities of BSE and IPA spectra over Raman shifts ranging from 900 cm$^{-1}$ to 1050 cm$^{-1}$, as a function of 
the excitation energy $\hbar \omega_L$. The $(\omega_L-\omega_R)^4$ prefactor, see Eq.~\ref{eq:fullsum0} has not been included, and the spectrum has 
been renormalized with respect to the value at $\hbar \omega_L=1.0~$eV. 
Note the logarithmic scale of the intensity.
The inset shows the ratio between integrated intensities obtained by the IPA and the BSE.}
 \label{fig:IBSE+RPA_ren}
\end{figure*}

\begin{figure}
 \centering
 \includegraphics[height=6.5cm]{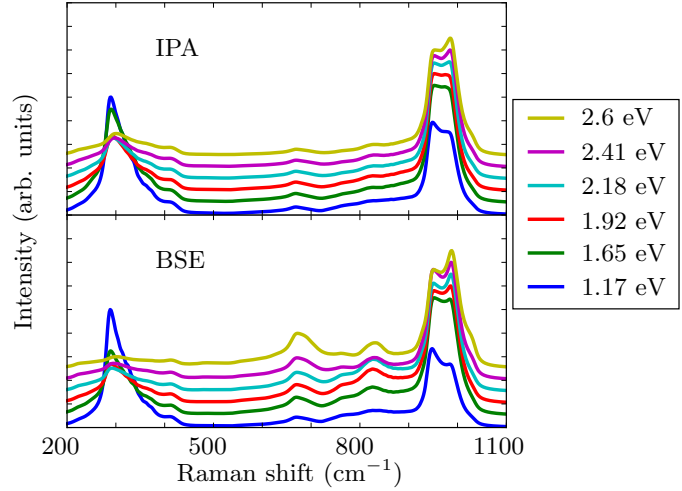} 
 \label{fig:raman_BSE}
 \caption{(Color online). Second-order Raman spectra calculated within the IPA 
(top) and the BSE (bottom) for several excitation energies $\hbar \omega_L$. Each spectrum is 
normalized with respect to its highest value, as in Fig.~\ref{fig:renucci}. 
 \label{fig:raman_BSE}
 }
\end{figure}

We first describe the {\it normalized} intensities for Raman shifts in the 900-1100 cm$^{-1}$ range. Such results are presented in the right panel of Fig.~\ref{fig:renucci}. 
On both levels of theory, IPA and BSE, the evolution of the intensity of the Raman shift with increasing light frequency (or photon excitation energy) 
follows very closely the experimental data. 
While most of the spectral weight is localized around 950 cm$^{-1}$ for 
the lowest excitation energy of 1.17 eV, the shoulder below 1000 cm$^{-1}$ is 
getting more pronounced with increasing excitation energy and starts to dominate 
above 2.4 eV. A third peak around 1030 cm$^{-1}$ that is hardly visible in the bottom 
curves, is particularly pronounced only for the highest excitation energy of 
3.41 eV. In excellent agreement with experiment, its intensity is nearly the 
same as the peak at 950 cm$^{-1}$, and somewhat lower than the one at 980 
cm$^{-1}$.

Second, we examine the {\it integrated} intensities. Fig.~\ref{fig:IBSE+RPA_ren} shows a comparison between the IPA and the BSE absolute intensity, 
divided by $(\omega_L - \omega_R)^4$ (see Methods section for more detail),
integrated over the range between 900 and 1100 cm$^{-1}$, as a function of the excitation energy.
Even if in this range of frequencies the relative intensities computed
in the two formalisms are very similar as shown in Fig.~\ref{fig:renucci}, 
their absolute intensities differ by a factor 5 for excitation energies around 3 eV, 
as exhibited in the inset of the Fig.~\ref{fig:IBSE+RPA_ren}. Clearly, excitonic effects
are important, which will be analyzed in the Discussion section.

The Raman spectrum in a broader range of Raman shifts is also obtained.
In addition to the enhancement of the absolute intensity, similar to the one shown in Fig.~\ref{fig:IBSE+RPA_ren},
Fig.~\ref{fig:raman_BSE} shows that the use of the BSE, that is, the inclusion of exciton-phonon coupling,
results in a pronounced intensity increase in the part of the Raman spectrum 
between 600 and 900 cm$^{-1}$, especially for excitation energies between 2 and 3~eV.
This additional excitonic effect will also be discussed later.

\begin{figure*}
 \centering
 \includegraphics[width=8.5cm]{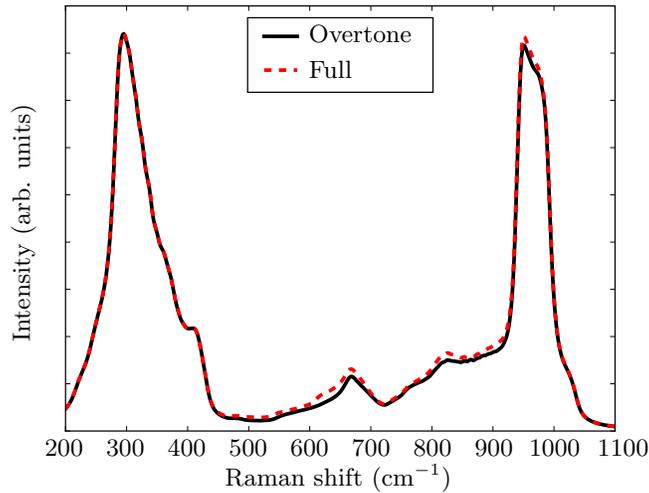}
 \caption{(Color online). Second-order Raman spectrum obtained by differentiating the dielectric constant (at $\hbar \omega_L=0$) obtained 
within density-functional perturbation theory. ``Full'' means that the emission of all possible phonon pairs are considered in the calculation, 
while for the ``overtone'' graph, only the emission of time-reversed phonon pairs are taken into account.}
 \label{fig:DFPTnonover}
\end{figure*}

In order to compute these results, we have made the approximation that the second-order
contribution to Raman spectrum is dominated by the emission of two phonons that are connected by time-reversal symmetry: same frequency, and time-reversed collective eigendisplacement and momentum.
From a group-theoretical point of 
view, derivatives with respect to such phonon pairs will always be 
active and contribute to the fully symmetric component of the 
spectrum~\cite{Cardona1975}. However, group-theoretical arguments
do not insure the vanishing of other contributions for arbitrary phonon wavevectors.
While this approximation is justified a posteriori by the agreement show in Fig.~\ref{fig:renucci},
it has been nevertheless possible to theoretically examine its validity, in the specific case of vanishing
excitation energy. 
We tested this approximation by computing the 
second-order derivatives of the dielectric constant obtained within 
Density-Functional Perturbation Theory (DFPT), where the electric field is used as a perturbation~\cite{Gonze1997a}. 
In this approach, the excitation energy vanishes, but unlike in the IPA, 
the local-field effects are taken into account, which makes its level of sophistication intermediate between IPA and BSE.
Fig.~\ref{fig:DFPTnonover} shows the dominance of the derivatives by the overtone pair of 
modes (by around one order of magnitude with respect to other contributions in the worst case), 
which fully justifies our approach.
This 'overtone' approximation was a crucial step in making the BSE computation tractable, 
as discussed in the Methods section, and to be able to treat excitonic effects. The confirmation
of the validity of the 'overtone' approximation for silicon, 
discussed in prior works, also for other simple semiconductors~\cite{Temple1973,Wang1973,Go1975,Renucci1975,Trommer1977,Cardona1982,Cardona1985,Grein1991,Garcia1994,Strauch1996} is thus another significant result of our work. However, there are several cases in which features can be assigned to combination modes, like in Refs.~\onlinecite{Pasternak1974,Reich1995,Venezuela2011}. In such case, the identification
of relevant combinations might be done in the static limit using DFPT, followed by the computation of the contributions from specific mode combinations
in the relevant energy spectrum (as is done already in our work, for degenerate phonon modes).

\section*{Discussion}

Our first-principles theory can provide insight into the origin of the features seen in Fig.~\ref{fig:renucci},
and then can shed light on the underlying physics, especially the role of excitonic effects. 
For this purpose, we first present the expression that yields the theoretical Raman intensity $I$
in Figs.~\ref{fig:renucci} (right part), ~\ref{fig:IBSE+RPA_ren}, and ~\ref{fig:raman_BSE}.
For a Raman shift $\hbar \omega_R$, at light excitation energy $\hbar \omega_L$, 
$I$ is as follows :
\begin{align}
I^{\alpha \beta}
(\omega_R,\omega_L) =\frac{\hbar^2(\omega_L - \omega_R)^4}{(4\pi)^2 c^4}
\frac{1}{N_{q}}
\sum_{m\textbf{q}} 
\left\{ 
\left[
\dfrac{\partial^2\varepsilon^{\alpha \beta}(\omega_L)}{\partial U^{m\textbf{q}} \partial U^{m(-\textbf{q})}} 
\frac{n(\omega_{m\textbf{q}}) + 1}{2 \omega_{m\textbf{q}}}
\right]^2 
\right. 
\delta(\omega_R-2 \omega_{m\textbf{q}}) 
\Bigg\}.
\label{eq:fullsum0}
\end{align}
The tensorial nature of $I$ ($\alpha$ and $\beta$ referring to the polarization of incoming and outgoing light) will not play an important role in our discussion.
Apart prefactors, the Raman intensity $I$ is essentially obtained by summing over contributions from all phonons, 
characterized by their momentum $\textbf{q}$ and branch index $m$. 
Such contributions are products of the {\it change
of the dielectric response $\bf{\varepsilon}$} due to each phonon and its time-reversed homolog, by
a function that depends {\it only on the phonon frequency $\omega_{m\textbf{q}}$} and a Dirac-delta, 
that expresses the conservation of energy.
This formula is explained in more detail in the Methods section,
which provides its derivation, including the ``overtone'' approximation,
as well as the description of dielectric response calculations, with or without excitonic effects.

In Eq.~\eqref{eq:fullsum0}, the Dirac-delta function, integrand of the two-phonon 
density of states
\begin{equation}
 g_{2DOS}(\omega_R)=\frac{1}{N_q} 
\sum_{m\textbf{q}} 
\delta(\omega_R-2 \omega_{m\textbf{q}})
\label{eq:phdos0}
\end{equation} 
is actually weighted by a function of the phonon mode and excitation energy (the term in square brackets in Eq.~\eqref{eq:fullsum0}).~\cite{Cardona1982,Weber2000}
One can expect such weighting factor 
to be a reasonably smooth function of 
the wavevector $\textbf{q}$. By contrast, the 
Dirac-delta function of Eq.~\ref{eq:phdos0} 
needs to be properly integrated throughout the Brillouin zone. 
It induces van Hove singularities, that dominate the overall shape of $I$. 

\begin{figure}
 \centering
 \includegraphics[width=15cm]{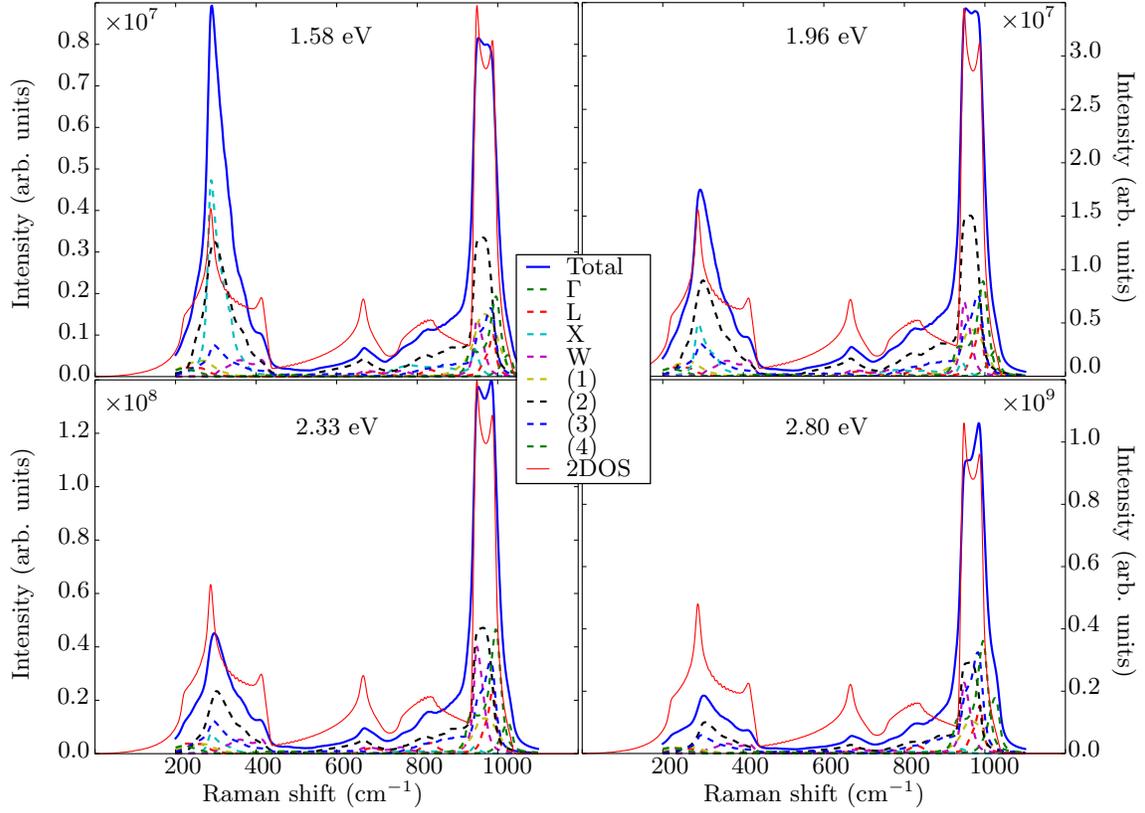}
 \caption{(Color online). Diagonal component of the second-order Raman 
spectrum obtained in the IPA (blue lines), and 
its wavevector decomposition for four different
excitation energies $\hbar \omega_L$: 1.58 eV, 1.96 eV, 2.33 eV and 2.80 eV.
Eight $\textbf{q}$ wavevectors are considered, in reduced coordinates : 
$\Gamma~=(0,0,0)$, 
L $=(1/2,1/2,1/2)$,
X $=(1/2,1/2,0)$,  
W $=(1/4,1/2,3/4)$, 
(1) $=(1/4,1/4,0)$, (2) $=(1/2,1/4,0)$, (3) $=(-1/4,1/4,0)$, 
(4) $=(1/4,1/4,1/4)$. The two-phonon density of states is also 
represented for sake of comparison (red lines).\label{fig:evolq}}
 \label{fig:decompRaman}
\end{figure}

\begin{figure}[th]
 \centering
 \includegraphics[width=10cm]{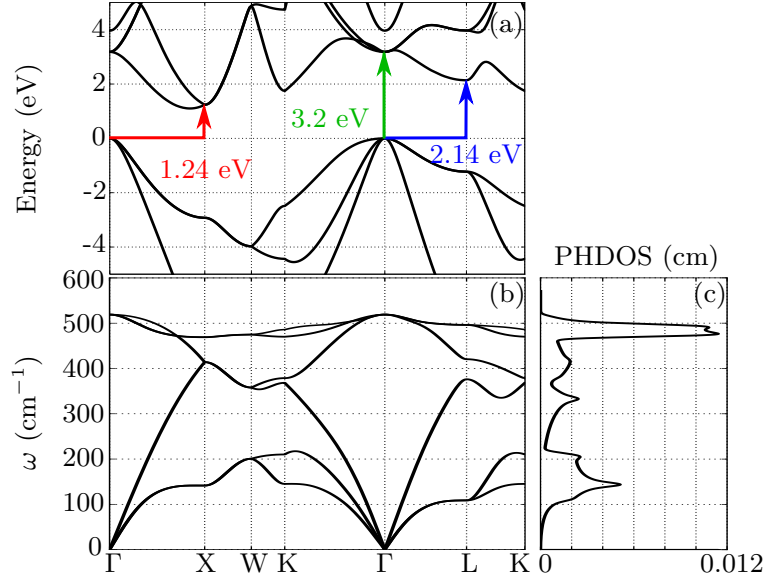}
 \caption{(Color online). (a) Electronic Kohn-Sham band structure 
of silicon, with a scissor shift of 0.65 eV for the conduction 
bands. 
Possible direct transitions at $\Gamma$ and indirect transitions from $\Gamma$ 
to X and from $\Gamma$ to L are indicated by the green, red, and blue arrows, 
respectively. (b) Phonon band structure and (c) phonon density of states.}
 \label{fig:bs}
\end{figure}

\begin{figure}[h]
 \centering
 \includegraphics[width=10cm]{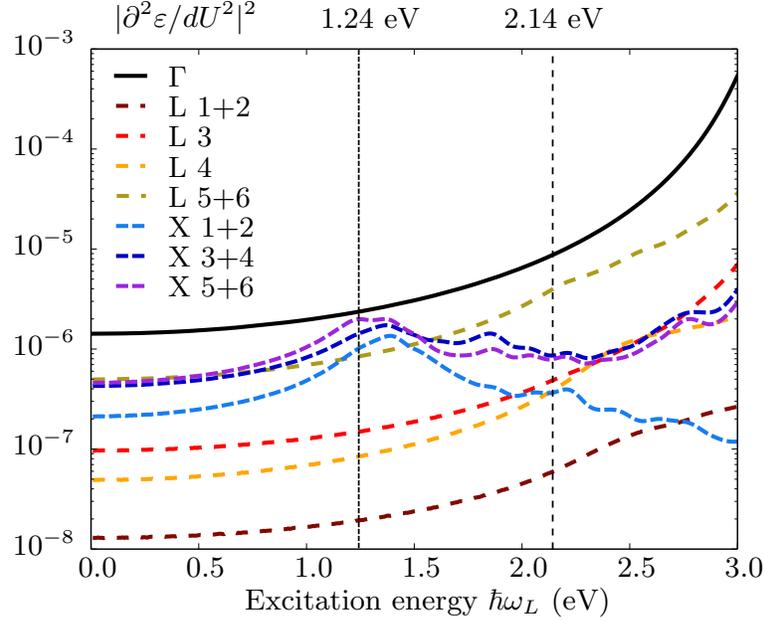}
 \caption{(Color online). Second derivatives of the IPA dielectric function 
with respect to the excitation energy for different modes at the $\Gamma$, $X$, 
and $L$ points. The notation ``$\textbf{q}$ m'' represents the $m^\text{th}$ 
branch of a q-point ($\Gamma$, $X$, or $L$), while ``q $m+n$'' represents 
the sum of the corresponding derivatives of the degenerate branches $m$ and 
$n$ at this point. The electronic transition thresholds for $X$ (1.24 eV) and 
$L$ (2.14 eV) are also indicated.}
 \label{fig:d2eps}
\end{figure}

The second-order Raman scattering intensity can thus be decomposed into contributions 
coming from different points of the Brillouin zone for phonons. The assignment
of peaks or characteristic features to selected phonons for Si had been established already in the seventies~\cite{Temple1973,Wang1973}.
Our methodology now allows one to examine accurately the behaviour of the intensity of each feature as a function of the
light excitation energy $\hbar \omega_L$, especially close to the resonance, 
as first attempted in the nineties by Grein and Cardona~\cite{Grein1991} 
using the theoretical and computational tools available at that time.
This is shown in Fig.~\ref{fig:evolq}, 
where the evolution of the diagonal component of the IPA Raman spectrum 
with respect to the excitation energy is depicted for eight characteristic wavevectors in the Brillouin zone.
In this framework, `optical' 
transitions between different electronic wavevectors, i.e., between $\textbf{k}$ and 
$\textbf{k}+\textbf{q}$ with non-zero $\textbf{q}$ are forbidden in the equilibrium structure. 
The phonon vibrations, however, may change some of these 
transitions from forbidden to allowed, giving rise to non-zero momentum matrix 
elements, thus leading to finite second-order derivatives of the dielectric 
function with respect to the phonon displacement. 
As a reference, the electronic and phononic band structures are shown in Fig.~\ref{fig:bs},
along with the one-phonon density of states (related to the two-phonon density of state given in Eq.~\eqref{eq:phdos0},
by a simple scaling of the phonon frequency axis).
As shown in Fig.~\ref{fig:bs} (b), at the $X$ point, the highest phonon 
frequency is about 470 cm$^{-1}$, twice this value being 940 cm$^{-1}$; the 
electronic resonance energy is 1.24 eV. At the $L$ point, twice the highest 
phonon frequency is slightly larger around  1000 cm$^{-1}$, with a resonance at 
2.14 eV. Likewise, for the $\Gamma$ point, the corresponding values are 1040 
cm$^{-1}$ and 3.2 eV. From a careful inspection of Fig.~\ref{fig:evolq},
we can conclude that the peak structures in Fig.~\ref{fig:renucci} are related to 
these resonances. To highlight their behavior as a function 
of the excitation energy, we plot in Fig.~\ref{fig:d2eps} the second 
derivatives of 
the dielectric function at various points of the BZ. At the $X$ point, it shows 
a clear maximum at the energy of the indirect band gap. The four curves related 
to the $L$ point do not peak at their resonance frequency, but increase with 
excitation energy, with a reduced rate above 2 eV. The 
curve at $\Gamma$ grows unabated even beyond 3.0 eV. From such different 
characteristics, one can also conclude that the two-phonon density of states 
alone is 
not sufficient to describe second-order Raman spectra in general, and 
even more so in resonance condition.


\begin{figure}[h]
 \centering
 \includegraphics[width=8cm]{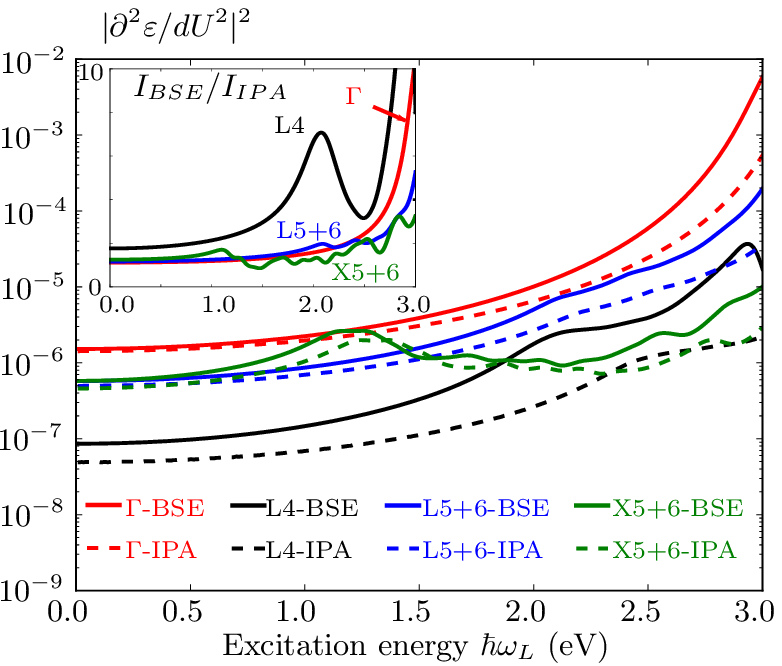}
 \caption{(Color online). Second derivatives of the dielectric function with respect to the excitation 
 energy for the highest-energy modes at $\Gamma$, $X$ and $L$ as well as the fourth mode at the 
 $L$ point. The inset shows the ratio between the BSE and the IPA results.}
 \label{fig:d2eps_BSEv2}
\end{figure}

To assess the impact of excitonic effects, we have performed the 
corresponding calculations with the derivatives of the dielectric function for 
phonon wavevectors $\Gamma$, $X$, and $L$ obtained by solving the BSE. A 
comparison with the IPA results is given in Fig.~\ref{fig:d2eps_BSEv2}, where 
we emphasize the logarithmic scale of the vertical axis. The corresponding Raman 
spectra obtained for several excitation energies are 
presented in Fig.~\ref{fig:raman_BSE}. As a general trend, the ratio between the 
BSE and IPA intensities (highlighted in the inset) increases with the excitation energy, 
exhibiting a maximum at the direct band gap. This effect is similar 
to the first-order 
enhancement effect reported in Ref.~\onlinecite{Gillet2013}. A very 
strong excitonic effect can be observed between 2 and 3 eV for the third and 
fourth mode of the $L$ point, as depicted in Fig.~\ref{fig:d2eps_BSEv2} for the 
latter. It can be explained by the presence of bound exciton states (inside the gap) with finite 
momentum \cite{Gatti2013} that become visible at second 
order due to the 
presence of finite-momentum phonons. The ratio between BSE and IPA results 
dramatically increases in this energy range due to the changed onset of the 
second-order response. The other $L$-point modes, around 100 and 500 cm$^{-1}$, 
do not exhibit strong coupling with excitons. Overall, exciton-phonon coupling 
results in a pronounced intensity increase in the part of the Raman spectrum 
which is dominated by $L$ modes, for excitation energies between 2 and 3~eV and 
Raman shifts between 600 and 900 cm$^{-1}$ (Fig.~\ref{fig:raman_BSE}).
This result is a central outcome of the present work.

For the other modes, indirect transitions appear less important for the 
resonance behavior than direct transitions. For example, in the regions around 
1000 and 1200~$\text{cm}^{-1}$, all the intensities increase with energy with a 
similar rate. Since the relative intensities are well reproduced within the IPA, 
this explains why the IPA spectra in this range are in already quite good agreement with 
experiment (see Fig.~\ref{fig:renucci} for a comparison between IPA and BSE 
in this range). On the other hand, an accurate description of absolute 
intensities requires the inclusion of excitonic effects as they lead to changes 
of one order of magnitude for excitation energies of around 3 eV, as already shown in Fig.~\ref{fig:IBSE+RPA_ren}.

Silicon, a paradigmatic material in solid-state physics, might however not be the material
in which exciton-phonon coupling has the biggest impact on the second-order Raman spectrum.
Indeed, for materials that possess infra-red active phonon modes (unlike silicon), the Fr\"{o}hlich coupling with
electrons diverges for small wavevectors, and dominates the exciton-phonon interaction. 
This has been explored for the second-order Raman case on the basis of model Hamiltonians in the eighties and nineties, see e.g. Ref.~\onlinecite{Garcia1994} and references therein.
The present approach can be generalized to the treatment of such Fr\"{o}hlich coupling.~\cite{Verdi2015} 

\section*{Conclusions}
Thanks to a new method that combines frozen-phonon 
supercell calculations of dielectric functions and their derivatives with 
phonon spectra from density-functional perturbation theory, 
we have been able to compute 
frequency-dependent second-order Raman spectra. We have applied our 
approach to the computation of the second-order Raman intensities of 
silicon for a 
series of excitation energies. We not only reach excellent agreement with 
measured spectra \cite{Renucci1975} but provide insight into the resonance 
behavior through analysis in terms of electronic structure and contributions 
from phonons at different points of the Brillouin zone. Excitonic effects turn 
out to be crucial to determine absolute Raman intensities as already found for 
the  first-order scattering \cite{Gillet2013}. Moreover, we have shown that strong 
excitonic effects may also appear selectively for some 
vibrational modes of different phonon wavevectors, modifying 
the relative intensities of different peaks. Our approach leads to a deeper 
understanding of the physics involved in second-order Raman 
processes and facilitates the interpretation of experimental 
results. While in this article we have 
focused on silicon---a simple but prototypical material---our methodology is generally 
applicable to a broad range of materials.

\section*{Methods}

We focus first on the theoretical methods, and then we give the 
computational details.

In the harmonic approximation for the vibrational properties of periodic solids, the following expression can
be used to compute the Raman spectrum ~\cite{Cardona1985,Knoll1995} for two 
Stokes processes:
\begin{align}
 I^{\alpha \beta}(\omega_R,\omega_L) =\frac{(\omega_L - 
\omega_R)^4 \hbar^2}{(4\pi)^2 
c^4}
\frac{1}{N_{q}}
\sum_{\textbf{q}} 
\sum_{m_1,m_2} 
\left\{ 
   \left[
   \dfrac{\partial^2\varepsilon^{\alpha 
   \beta}(\omega_L)}{\partial U^{m_1\textbf{q}} \partial U^{m_2(-\textbf{q})}} 
   \right]^2 
\frac{1}{4 \omega_{m_1\textbf{q}} \omega_{m_2(-\textbf{q})}} \right. \nonumber \\
\left. [n_{m_1\textbf{q}} + 1] [ n_{m_2(-\textbf{q})} + 1 ] 
\delta_{\gamma}\textbf{(} \omega_R-(\omega_{m_1\textbf{q}}+\omega_{m_2(-\textbf{q})}) \textbf{)} \right.\Bigg\},
\label{eq:fullsum1}
\end{align}
where $\alpha$ and $\beta$ are Cartesian indices related to the polarization of the 
incoming and outgoing light, 
 $\textbf{q}$ are phonon wavevectors sampling the full BZ homogeneously 
(with a total of $N_q$ points), $m_1$ 
and $m_2$ are phonon branch indices, $\omega_R$ is the spectral 
frequency (Raman shift), $\omega_L$ is the excitation frequency, $c$ the speed of 
light, $\omega_{m\textbf{q}}$ is the phonon 
frequency of the $m^{\text{th}}$ phonon branch at $\textbf{q}$, $n_{m\textbf{q}}$ is the 
temperature-dependent phonon occupation factor
$n_{m\textbf{q}} = {(e^{\frac{\hbar \omega_{m\textbf{q}}}{kT}}-1)}^{-1}$
and $\delta_{\gamma}$  is a Dirac-delta 
function replaced
by a Lorentzian function of width $\gamma$ for numerical purposes. In Eq.~\eqref{eq:fullsum1} , 
the derivative of the dielectric function, $\varepsilon$, is obtained 
through an expansion of the dielectric function with respect to the phonon 
eigendisplacements.
Extracting the second-order process from this expansion and using the adiabatic 
approximation~\cite{Knoll1995} gives, for the electronic part, the following second-order
derivative:
\begin{align}
 &\dfrac{\partial^2\varepsilon^{\alpha 
\beta}}{\partial U^{m_1\textbf{q}} \partial U^{m_2(-\textbf{q})}}
= \sum_{\kappa 
\alpha l, \kappa' \beta l'} \dfrac{\partial^2\varepsilon^{\alpha 
\beta}}{\partial R^l_{\kappa \alpha} \partial R^{l'}_{\kappa' 
\beta}} e^{i \textbf{q} (\textbf{R}_l - \textbf{R}_{l'})} U^{m_1\textbf{q}}_{\kappa 
\alpha} U^{m_2(-\textbf{q})}_{\kappa' 
\beta},
\end{align}
where $\textbf{R}_l$ is the coordinate of the $l^{\text{th}}$ periodic cell, 
$R^{l}_{\kappa \alpha}$ the coordinate of the $\kappa^{\text{th}}$ atom along 
$\alpha$ axis in the $l^{\text{th}}$ periodic cell. $U^{m_\textbf{q}}_{\kappa 
\alpha}$ are the atomic eigendisplacements of the dynamical 
problem, following the conventions of Ref.~\onlinecite{Gonze1997a}. 

The numerical evaluation of Eq.~\ref{eq:fullsum1} is highly challenging
since it requires careful integration over the Brillouin zone and summation over
two mode indices. This contrasts with the summation over only one mode index, at the zone center,
required for first-order Raman intensities.

We now describe how to numerically evaluate from first-principles the above-mentioned equations.
A full ab-initio phonon band structure can be obtained from DFPT~\cite{Baroni2001,Gonze2005a}. 
As a first step, the phonon frequencies $\omega_{m\textbf{q}}$ and the 
phonon eigendisplacements $U^{m\textbf{q}}_{\kappa \alpha}$ are computed for 
a set of homogeneous $\textbf{q}$-points, and interpolated 
on a denser mesh of wavevectors thanks to the knowledge of interatomic force constants, as described in 
Ref.~\onlinecite{Gonze1997a}. 
Unfortunately, existing DFPT implementations do not provide second derivatives 
of the frequency-dependent dielectric function with respect to atomic positions especially when excitonic 
effects are to be included. We thus numerically differentiate the dielectric 
function with respect to the eigendisplacements, which requires supercells of 
different size, depending on the phonon wavelength. 
In the frozen-phonon approach, a 
supercell,
compatible with the selected wavevector $\textbf{q}$ is generated: the axes of this 
supercell $\left\{ \textbf{R}'_j \right\}$ fulfill the constraint $e^{i 
\textbf{q} \textbf{R}'_j} = 
1$, which means that $\textbf{q}$ is a reciprocal vector of the 
supercell. The atoms of the supercell are then displaced with real displacements 
defined 
from the phonon eigendisplacements and phase factors
\begin{align}
 u_{\kappa \alpha l} \triangleq \frac{1}{2} \left[ U_{\kappa 
\alpha}(\textbf{q}) e^{i \textbf{q} \textbf{R}_l} + U_{\kappa 
\alpha}(-\textbf{q}) e^{-i 
\textbf{q} \textbf{R}_l} \right], ~~~~~ v_{\kappa \alpha l} \triangleq \frac{-i}{2} \left[ U_{\kappa 
\alpha}(\textbf{q}) e^{i \textbf{q} \textbf{R}_l} - U_{\kappa 
\alpha}(-\textbf{q}) e^{-i \textbf{q} \textbf{R}_l} \right].
\end{align}
In the supercell, $\textbf{k}+\textbf{q}$ points of the Brillouin zone are folded back 
to $\textbf{k}$ points of the 
supercell. This allows for probing indirect transitions.

The calculations of the optical spectrum can be 
performed with different levels of accuracy, ranging from the independent 
particle approximation to time-dependent DFT (TDDFT) and, finally, 
many-body perturbation theory (MBPT)
with the inclusion of excitonic effects by means of the Bethe-Salpeter 
equation~\cite{Hanke1979,Onida2002}. 

We now detail why even for a small system like silicon such calculations are 
rather demanding (extremely demanding in the BSE case).
First, for any wavevector, the summations over the $N_{modes}$ phonon modes in Eq.~\ref{eq:fullsum1} require $N_{modes} \times N_{modes}$
evaluations of the derivatives. 
If the dielectric tensor is evaluated on a grid of $5 \times 5$ pairs of displacements,
in order to get the second-order derivatives,
the number of calculations required for a single wavevector in the BZ becomes $25 \times 6^2 = 900$ evaluations of the dielectric tensor for silicon.
In contrast, for first-order calculations, the number of computations for the zone-center wavevector amounts to 5 evaluations for each mode, thus $5 \times 6 = 30$
evaluations of the dielectric function to reach the final Raman spectrum of silicon.
This number can be even reduced to only 5 by using symmetry arguments: acoustic modes
do not contribute and the three other modes are degenerate.

It should also be noted that, unlike first-order spectra, 
second-order resonant Raman scattering requires the use of supercells 
commensurate with the wavevectors, and the computational cost increases 
quickly with the size of the supercell. Indeed, the number of
bands included in the calculations scales linearly with the number of atoms. 
In the case of IPA, the number of electronic transitions scales linearly with the product of the number of valence and conduction bands and linearly
with the number of k-points.
Taking into account the folding of bands in the case of supercell, 
and the computational cost required for the dipole matrix elements, 
the overall scaling is roughly cubic with the size of the supercell.
In BSE, however, the kernel introduces coupling between transitions,
and the number of matrix elements in the transition
basis set scales with the square of the product of the number of valence and conduction bands, and with
the square of the number of k-points. 
Because of the reduction of the number of k-points with larger supercells, 
the number of matrix elements therefore increases with the square of the size of the supercell.
Since the evaluation of the screened Coulomb matrix elements requires a 
double sum over planewaves, we conclude that the ratio between a BSE run 
for a supercell and the one for a unit cell increases as the fourth power of the number of replicas in the supercell. Finally,
as already observed in Ref.~\onlinecite{Gillet2013}, the derivatives
of the dielectric function are highly sensitive to the sampling of the BZ, leading to
very demanding calculations in the case of BSE.
Even if the frozen-phonon methodology requires a large 
amount of evaluations of dielectric functions, we have 
succeeded to apply the technique to three-dimensional materials by using 
approximations and numerical techniques as follows.

As mentioned in the Results section, the so-called ``overtone'' contributions, i.e., 
considering only derivatives coming from twice the same phonon mode, except when 
degenerate, should largely dominate other second-order contributions. 
In Eq.~\eqref{eq:fullsum1} this corresponds to neglecting all 
terms where $\omega_{m_1} \ne \omega_{m_2}$ and yields Eq.~\eqref{eq:fullsum0}.

The overtone approximation allowed us to reduce significantly the number of required 
computations. Eventually, only one-dimensional derivatives are needed for 
non-degenerate modes, reducing the number of evaluations required for one wavevector
from 900 for the full expression
to $5 \times 6 = 30$. For high-symmetry points of the BZ, few 
two-dimensional 
derivatives are still required between two different modes belonging to the 
same degenerate subspace. As shown in Fig.~\ref{fig:renucci}(b), the frequency-dependent spectrum is well reproduced for  silicon with this approximation. This figure corresponds to the $Z(XX)\overline{Z}$ scattering configuration.

In order to further reduce the computational load, the second-order 
derivatives of the dielectric function have been evaluated only on a coarse 
mesh of phonons wavevectors, while the phonon frequencies, the occupation numbers 
and  Dirac functions have been calculated via DFPT techniques and summed on a dense 
mesh. This is equivalent to taking an averaged density of states around the 
coarse points as the density function of Ref.~\onlinecite{Knoll1995}.

The corresponding grid parameters are now described, as well as other computational details.
We use the ABINIT 
package~\cite{Gonze2009,Gonze2016}, with Troullier-Martin 
pseudopotentials (available from the ABINIT package\cite{AbinitWebsite}). Relaxed cell 
parameters of 10.20 Bohr are used with 
cut-off values for the wavefunctions of 8 Hartree. The DFPT phonons 
displacements 
and frequencies are interpolated from a coarse $\Gamma$-centered 
phonon grid of $8 \times 8 \times 8$ to a non-shifted $70 \times 70 \times 
70$ mesh, to obtain the phonon density of states, following Ref.
 ~\onlinecite{Gonze1997a}, while a 4-times shifted $8 \times 8 \times 8$ sampling
 is used for the electrons.

When simulated in the IPA (without local-fields effects)~\cite{Ambrosch2006}, the dielectric 
function 
is evaluated for different supercells corresponding 
to a non-shifted $4 \times 4\times 4$ phonon wavevector grid 
(containing 64 wavevectors), while 
for the electronic states, the equivalent of a centered $24 \times 
24 
\times 24$ grid
for the primitive cell is used. In the supercell case,
the Brillouin zone is folded, and the integration grid adapted 
accordingly. The convergence has been checked on the final Raman spectrum with 
respect to a 4-times shifted $2 \times 2 \times 2$  centered phonon wavevector 
grid (containing 32 wavevectors).
A scissors operator of 0.65 eV is applied on 
top of the Kohn-Sham eigenvalues to mimic the opening of the gap by GW. 

In order to assess the importance of 
excitonic effects, we have performed BSE calculations of the derivatives of the 
dielectric function for phonons with $\Gamma$, $X$ and $L$ wavevectors 
(corresponding to a $2 \times 2 \times 2$ $\Gamma$-centered grid). Dielectric 
functions obtained on $12 \times 12 \times 12$ wavevectors in the BZ have been 
averaged following the technique described in Ref.~\onlinecite{Gillet2013} to reach 
samplings equivalent to $24 \times 24 \times 24$. 3 valence bands and 6 
conduction bands were used for the primitive cells, their numbers being 
increased proportionally with the number of atoms in the supercells. The model 
dielectric function described in Ref.~\onlinecite{Cappellini1993} with 
$\varepsilon^{\infty} = 12$ has been used to calculate the screening matrix 
elements. A cut-off energy of 4~Ha is used to describe the screening matrix in 
reciprocal space.

In order to present a Raman spectrum obtained with a 
sufficient number of phonon frequencies in the BSE framework, we have interpolated the ratio between 
BSE and IPA intensities towards a non-shifted $4\times4\times4$ k-point grid 
with a trilinear interpolation in reduced coordinates and used these ratios to correct
the IPA intensities.

When integrated intensities are presented, the $(\omega_L - \omega_R)^4$ factor is discarded.
This factor is present also in the Raman spectrum of reference wide-band gap materials, 
like $CaF_2$, to which the comparison is made in order to establish the absolute intensity. 
Hence, it can be discarded. Other factors are expected to be constant in wide band-gap materials.

\section*{Acknowledgments}

Y.G. and M.G. acknowledge financial support by the Fonds National de la 
Recherche Scientifique (FNRS, Belgium) and the ESF through the ESF/Psi-k2 
programme. C.D. and S.K. appreciate support from the German Science Foundation 
(DFG) through CRC 658. We thank Yann Pouillon and Jean-Michel Beuken for their 
valuable technical support and help with the test and the build system of 
ABINIT. Computational resources have been provided by the supercomputing 
facilities of the Universit\'e catholique de Louvain (CISM/UCL) and the 
Consortium des Equipements de Calcul Intensif en F\'ed\'eration Wallonie 
Bruxelles (CECI) funded by the Fonds de la Recherche Scientifique de Belgique 
(FRS-FNRS). The present research benefited from computational resources made 
available on the Tier-1 supercomputer of the F\'ed\'eration Wallonie-Bruxelles, 
infrastructure funded by the Walloon Region under the grant agreement 
n$^{o}$1117545 and the compute cluster DUNE at the Humboldt-Universit\"at of Berlin,
funded by the Einstein Foundation 
Berlin.

\section*{Author contributions}
Y.G. and S.K. developed the methodology.
Y.G. performed the calculations.
Y.G., C. D., X. G. analyzed the data.
S.K., M.G., C.D., X.G. supervized the work.
Y.G., M.G., C. D., X. G. wrote the paper.

\section*{Competing financial interests}
The authors declare no competing financial interests.

\end{document}